\documentclass[10pt,twocolumn,letterpaper]{article}

\usepackage[pagenumbers]{cvpr} 

\usepackage{amsmath}
\usepackage{tikz}
\usepackage{pgfplots}
\usepackage{floatrow}
\pgfplotsset{compat=1.18}
\usepackage{algorithm}
\usepackage{algpseudocode}

\definecolor{cvprblue}{rgb}{0.21,0.49,0.74}
\usepackage[pagebackref,breaklinks,colorlinks,allcolors=cvprblue]{hyperref}

\def\confName{CVPR}
\def\confYear{2025}

\definecolor{gold}{HTML}{008B8B}
\definecolor{grey}{HTML}{A9A9A9}

\title{FlipSketch: Flipping Static Drawings to Text-Guided Sketch Animations\vspace{-5mm}}

\author{
\href{https://hmrishavbandy.github.io/}{Hmrishav Bandyopadhyay} \hspace{.2cm} \href{https://personalpages.surrey.ac.uk/y.song/}{Yi-Zhe Song} \\
SketchX, CVSSP, University of Surrey, United Kingdom.  \\
{\tt\small \{h.bandyopadhyay, y.song\}@surrey.ac.uk}
}

\vspace{-0.4cm}

\begin{document}

\twocolumn[{%
\renewcommand\twocolumn[1][]{#1}%
\maketitle
    \captionsetup{type=figure}
    \vspace{-1cm}
    \includegraphics[trim={0 0 1.5cm 0}, width=\textwidth, scale = 0.1]{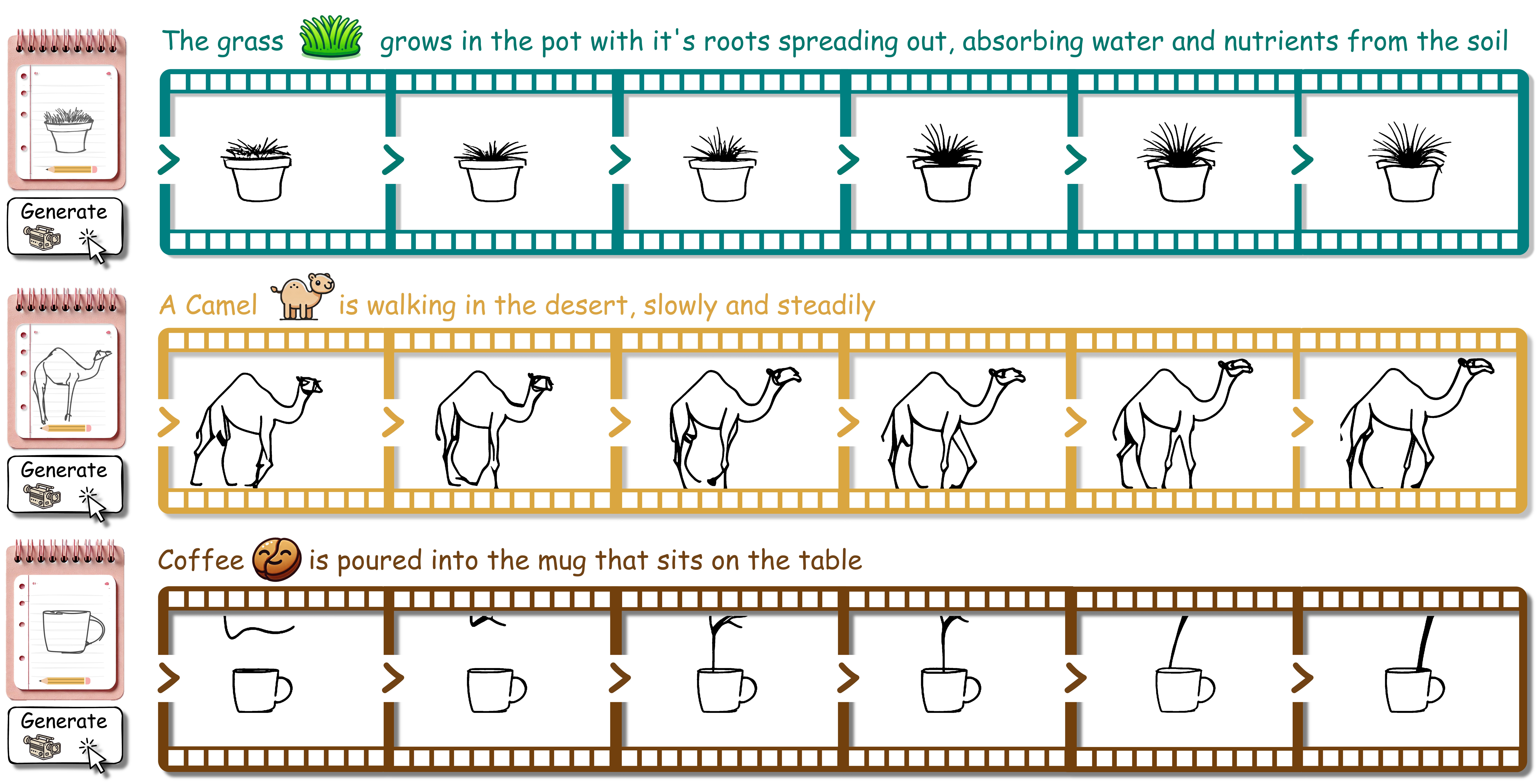} \\[-0.7cm]   
    \caption{We animate raster sketches using motion priors from pre-trained text-to-video diffusion models. Our generated animations are dynamic, with scene-level interaction that is not possible with stroke-constrained vector animation algorithms \cite{gal2024breathing, sketchvideosynthesis}. We control animations with input sketches, where generated videos closely preserve sketch identity without losing range of motion (GIFs in Suppl.)}
    \label{fig:teaser}
    \vspace{0.2cm}
}]

\begin{abstract}
\vspace{-1.2cm} \\

Sketch animations offer a powerful medium for visual storytelling, from simple flip-book doodles to professional studio productions. While traditional animation requires teams of skilled artists to draw key frames and in-between frames, existing automation attempts still demand significant artistic effort through precise motion paths or keyframe specification. We present FlipSketch, a system that brings back the magic of flip-book animation -- just draw your idea and describe how you want it to move! Our approach harnesses motion priors from text-to-video diffusion models, adapting them to generate sketch animations through three key innovations: (i) fine-tuning for sketch-style frame generation, (ii) a reference frame mechanism that preserves visual integrity of input sketch through noise refinement, and (iii) a dual-attention composition that enables fluid motion without losing visual consistency. Unlike constrained vector animations, our raster frames support dynamic sketch transformations, capturing the expressive freedom of traditional animation. The result is an intuitive system that makes sketch animation as simple as doodling and describing, while maintaining the artistic essence of hand-drawn animation.

\end{abstract}
\vspace{-0.7cm}

\section{Introduction}
\label{sec:intro}

Remember those tiny stick figures dancing in the corners of your textbook pages? That simple flip-book magic captured something fundamental -- our desire to breathe life into our drawings! We present a system that brings back that flip-book magic -- just doodle your idea and describe how you want it to move.

From these playful beginnings, traditional animation studios like Disney perfected the art of bringing drawings to life through a process where lead animators draw key frames and skilled artists fill in the intermediate frames -- a technique known as in-betweening. This workflow remains the industry standard today, requiring teams of skilled artists to manually craft each frame. Existing attempts at automating this process \cite{brodt2024inbetweening} still demand significant artistic effort -- requiring users to specify precise motion paths \cite{thorne2004motion}, control points, or multiple keyframes \cite{yang2017context}, much like a digital puppet show rather than true animation.

Recent vector-based techniques \cite{gal2024breathing}, while innovative, remain limited by their stroke-by-stroke manipulation \cite{vinker2022clipasso} approach, inherently restricting the fluidity and expressiveness of the resulting animations. Such coordinate transformation methods miss the artistic essence of traditional animation -- the ability to freely redraw and reinterpret the subject across frames.

The transformation of static sketches into fluid animations presents three significant technical challenges: (i) adapting video generation models to maintain the distinctive aesthetic of hand-drawn sketches, (ii) ensuring temporal consistency by preserving the visual integrity of the input sketch across frames, and (iii) supporting unconstrained motion that enables dynamic modifications while maintaining visual coherence.

Our framework addresses these challenges through three technical innovations. First, we fine-tune a text-to-video diffusion model \cite{Wang2023ModelScopeTT} on synthetic sketch animations, enabling it to generate coherent line-drawing sequences while leveraging sophisticated motion priors from video data. Second, we introduce a reference frame mechanism based on DDIM inversion of the input sketch, extracting a canonical noise pattern that captures the sketch's essential style. Through iterative refinement of subsequent frames, we maintain fine-grained sketch details while allowing for natural motion evolution. Third, we develop a novel dual-attention composition during the denoising process, selectively transferring both coarse and fine-grained information across frames. This enables precise control over both identity preservation and motion fidelity in generated animations (see \cref{fig:teaser}).

In summary, our contributions are: (i) FlipSketch, the first system to generate unconstrained sketch animations from single drawings through text guidance, powered by motion priors from T2V diffusion models. (ii) A novel reference frame technique using iterative noise refinement that preserves sketch visual integrity across frames, akin to traditional animation principles. (iii) A dual-attention composition mechanism that enables fluid motion while preserving sketch identity, supporting dynamic transformations beyond simple stroke manipulation. Together, these advances make sketch animation as simple as doodling and describing, while maintaining the expressive freedom of traditional animation.

\section{Related Works}
\label{sec:related_works}

\subsection{Diffusion Powered Video Generation} Generation of open-domain videos from text prompts has seen rapid growth piggybacking off high fidelity text-to-image (T2I) diffusion frameworks \cite{rombach2022high, song2020score, ho2020denoising}. Popular text-to-video (T2V) models \cite{xing2023dynamicrafter, Wang2023ModelScopeTT} can generate complex motion with dynamic backgrounds, previously unseen with language-like sequence generation approaches \cite{tian2021good, denton2018stochastic}. Many of these T2V models \cite{bar2024lumiere} borrow heavily from the success of T2I frameworks \cite{saharia2022photorealistic}, often directly using pre-trained T2I as part of the video generation pipeline. This adaptation generally involves using frozen T2I for spatial generation within each individual frame and training new units for maintaining temporal consistency across video frames. Besides T2V adaptation, T2I models are also used for Image to Video (I2V) \cite{blattmann2023stable, zhang2023i2vgen, xing2023dynamicrafter} generation, where CLIP \cite{radford2021learning} features from input image guides frame denoising. Recent works \cite{yuan2024instructvideo} introduce fine-tuning approaches \cite{hu2021lora, black2023training} for T2V models to improve generation quality through human feedback \cite{wu2023human}. In this work, we use a similar approach, fine-tuning a pre-trained T2V model \cite{Wang2023ModelScopeTT} for text to sketch animation generation. We further condition the animation generation on a user-provided sketch to gain spatial control over generated videos.

\subsection{Sketch for spatial control} 

Freehand sketches \cite{ha2017neural, eitz2012humans} are generally represented as vector diagrams through lists (or sets \cite{ashcroftmodelling}) of coordinates for poly-lines \cite{ha2017neural} and parametric curves \cite{das2020beziersketch, vinker2022clipasso}. These vectors can be rendered on a blank 2D (or 3D \cite{3doodle, qu2024wired}) canvas to construct a raster sketch as a line-drawing \cite{chan2022learning} of an object \cite{eitz2012humans} or scene \cite{vinker2023clipascene, chowdhury2022fs}. Generative modelling of freehand sketches primarily focuses on vector diagrams, using language models \cite{ha2017neural, wang2021sketchembednet, mo2021general} and implicit representations \cite{bandyopadhyay2024sketchinr, das2022sketchode} to generate vector coordinates. Both vector and raster representations of sketches offer intuitive communication of complex spatially rich ideas \cite{sain2025freeview, bhunia2022adaptive}, making them the de-facto choice to assist generative modelling of 2D \cite{zhang2023adding,koley2024s} and 3D \cite{bandyopadhyay2024doodle} environments. Recent works additionally explore sketch-like binary maps for controlling the generative modelling of videos \cite{chen2023control, zhang2023controlvideo}. In this paper, we explore the animation of sketches as raster images through fine-tuned T2V models. We further explore spatial control from these animations in generating real world videos.

\subsection{Sketch animation}
Animating hand-drawn sketches has long been a focus of research in computer graphics \cite{gal2024breathing}. Early works automate sketch animation using information from user-defined paths \cite{thorne2004motion} and partial sketches  \cite{xing2015autocomplete}. More recently, computationally assisted animations generate in-between frames for sketch vectors \cite{yang2017context,jiang2022stroke} and rasters \cite{siyao2023deep, xing2024tooncrafter}, reducing time and effort to animate significantly. 

Indirectly borrowing from videos, recent works use T2V priors to animate vector curves \cite{liu2024dynamic} by optimizing motion \cite{gal2024breathing} of bézier control points \cite{vinker2022clipasso}. This optimization is made possible with Score Distillation Sampling (SDS) \cite{poole2022dreamfusion} and differentiable rendering of vector strokes \cite{diffvg} and can yield high quality animations \cite{gal2024breathing} without any training. However, per-sample optimization of this form usually consumes an unrealistic amount of time and compute, reducing the feasibility of such approaches. In this work, we directly predict raster sketch animations with fine-tuned T2V models to reduce time taken by iterative optimisation, speeding up the animation pipeline by a large margin.

\vspace{-0.1cm}
\subsection{Inversion for GenAI}
Generative modelling of images \cite{karras2019style, song2020denoising} and image features \cite{rombach2022high} allows editing them by inverting \cite{alaluf2022hyperstyle} into their latent representation and searching for edit directions \cite{li2023w-plus-adapter}. Image inversion has been extensively studied in the context of GANs \cite{alaluf2022hyperstyle} for editing image content \cite{alaluf2022third} or style \cite{yang2022pastiche}. Recent works use DDIM \cite{song2020denoising} for inverting images through T2I diffusion frameworks \cite{rombach2022high} with null prompts \cite{mokady2023null}, allowing for accurate image reconstruction from inverted noise. TF-Icon \cite{lu2023tf} reduces noise in image reconstruction with special prompts and schedulers \cite{lu2022dpm}. In this work, we use DDIM inversion to invert input sketches in the T2V latent space. This helps us condition video generation from both the input sketch and the text prompt.

\begin{figure*}[!hbt]
    \centering
    \vspace{-0.3cm}
    \includegraphics[width=\linewidth]{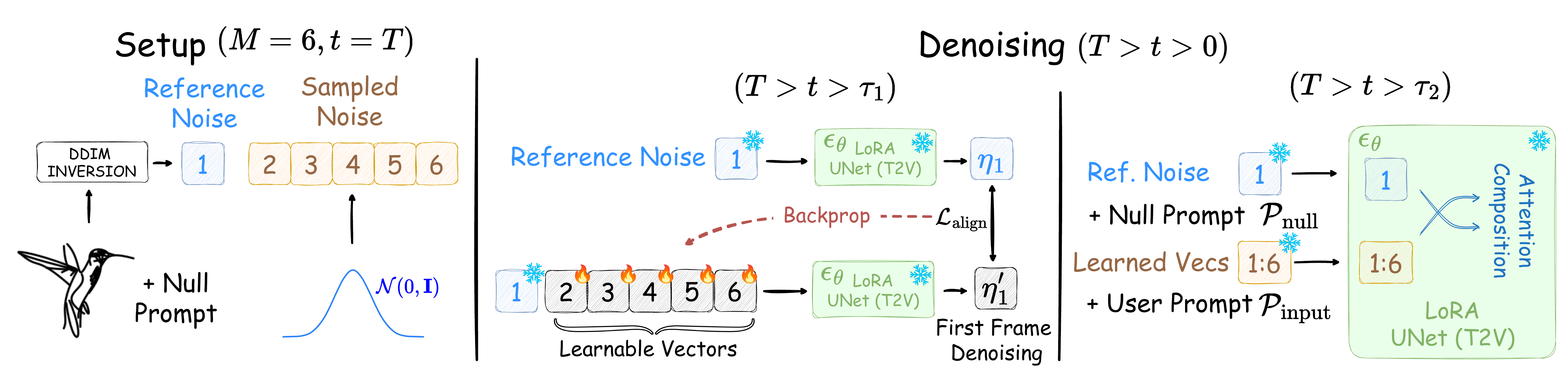}
    \vspace*{-0.8cm}
    \caption{Model Overview: (i) During setup, we invert the input sketch to act as the reference noise for the first frame, sampling from a standard normal for the rest. (ii) For timesteps within threshold $\tau_1$, we iteratively refine sampled noise for our reference noise (first frame) is denoised to the input sketch. (iii) We further compose attention maps for joint denoising of reference and sampled noise to influence all frames with first-frame information.\vspace{-0.6cm}}
    \label{fig:model_1}
\end{figure*}
\section{Background}
\vspace{-0.1cm}

\subsection{Low Rank Adaptation}
Low Rank Adaptations (LoRA) \cite{hu2021lora} of pre-trained large language models are trainable weights that can align these models to new tasks. These trainable weights are generally in the form of matrices $W^*$ added to frozen pre-trained parameters $W_0$ as $W_\text{LoRA} = W^* + W_0$. Following the hypothesis that pre-trained LLM weights have low intrinsic rank \cite{aghajanyan2020intrinsic}, the additional weights can be composed to have low rank as $W^*_{h_1\times h_2} = A_{h_1\times r} \times B_{r\times h_2}$, where $r << \text{min}(h_1,h_2)$. This allows for construction of $W^*$ in very few parameters ($h_1\times r + r\times h_2$) rather than learning the entire matrix $h_1\times h_2$. In practice, LoRAs of large models (like $\sim175$B parameter GPT2) use as less as $0.01\%$ learnable parameters \cite{hu2021lora} when fine-tuning to a new task. 

Recent works \cite{zanella2024low} extend LoRAs beyond language models to vision language frameworks like CLIP \cite{zanella2024low} and media generation with Stable Diffusion \cite{luo2023lcm, yuan2024instructvideo}. In this work, we adapt a popular text-to-video diffusion framework \cite{Wang2023ModelScopeTT} for sketch video generation with text and sketch guidance.

\vspace{-0.1cm}
\subsection{Diffusion for Video Generation}
\label{sec: bck_diff}
ModelScope T2V \cite{Wang2023ModelScopeTT} generates videos with diffusion in the low-dimensional latent-space \cite{rombach2022high} of a VQ-GAN\cite{esser2021taming}. Specifically, frames ($\in \mathbb{R}^{F\times H \times W \times 3}$) of videos are encoded as images to obtain their corresponding latents ($\in \mathbb{R}^{F\times \frac{H}{8} \times \frac{W}{8
}\times 3}$). These latents are noised and denoised with a 3D UNet consisting of spatial (within frame) and temporal (across frames) convolutional and attention layers. Spatial convolution and attention units treat frames as batches of images and perform denoising of individual frames. Temporal convolution and attention units capture correlation across frames to help with video consistency and coherence. We train a LoRA \cite{yuan2024instructvideo} on Modelscope T2V \cite{Wang2023ModelScopeTT} using synthetic animations from text prompts \cite{gal2024breathing} as training data.

\section{Proposed Methodology}
\textbf{Overview}:
Sketches are generally animated as sequences of vector frames, with individual strokes displaced at each frame \cite{gal2024breathing, sketchvideosynthesis} to convey motion. This localised displacement of vector strokes helps preserve structure (and identity) of non moving parts. However, the resulting animation is limited to displacing and scaling existing strokes as strokes can neither be added, nor removed. This significantly constrains animation possibilities, as a 2D sketch often represents only a partial view of a 3D object, requiring different strokes for different perspectives. 

To enable more flexible animations, we explore unconstrained raster sketches (as opposed to vectors) for generating sketch videos. Specifically, we represent animations as sequences of raster frames, relying on strong video diffusion priors of pre-trained T2V networks \cite{Wang2023ModelScopeTT} for consistent identity and content across frames. To adapt T2V frameworks for sketch-style video generation, we learn their Low Rank Adaptations with synthetic text-animation pairs. Next, we provide control over generated animations with user sketches, by modifying T2V attention units and noisy latents iteratively during denoising. We find that our approach allows us to preserve the identity of the input sketch in generated animations and reduces artefacts otherwise prominent with raster video generation pipelines.

\noindent \textbf{Baseline Text-to-Animation}: To generate animations from text prompts $\mathcal{P}_\text{input}$, we train a LoRA $\epsilon_\theta$ of pre-trained ModelScope T2V \cite{Wang2023ModelScopeTT} (backbone comparison in Suppl.) on synthetic vector sketch animations from \cite{gal2024breathing}.  
Our inference pipeline uses a text prompt $\mathcal{P}_\text{input}$ to iteratively denoise sampled noise $\{f^i_T\}_{i=1}^{M} \sim \mathcal{N}(0, \mathbf{I})$. At each timestep $t$ from $T\rightarrow0$, the denoising signal is obtained with network $\epsilon_\theta$ as: 
\begin{equation}
    \{\eta_i\}_{i=1}^{M}  = \epsilon_\theta(\{f^i_{t}\}_{i=1}^M, t, \mathcal{P}_\text{input})
\end{equation}
At $t=0$, $\{f^i_T\}_{i=1}^{M}$ is decoded with pre-trained T2V VQGAN \cite{esser2021taming} decoder to construct high resolution video frames. While the baseline T2V model can generate sketch animations, it often yields artefacts and watermarks from pre-training. We guide the generation with a user provided input sketch $ I_s\in \mathbb{R}^{H\times W \times 3}$ which significantly reduces artefacts. We finally post-process the output frames to constrain them to black strokes on a white canvas.

\subsection{Setup}
We encode the input sketch $ I_s\in \mathbb{R}^{H\times W \times 3}$ with pre-trained T2V VQGAN \cite{esser2021taming} and invert it with null-text inversion \cite{mokady2023null}. This involves performing DDIM inversion with null prompts $\mathcal{P}_\text{null}$ to obtain a noise vector at $t=T$ that can be denoised to $t=0$ to reconstruct exact input $I_s$ accurately. We use the inversion noise at $t=T$ as the \textbf{reference noise} $x^r_{T} \in \mathbb{R}^{\frac{H}{8}\times \frac{W}{8} \times 4}$ for generating the first frame, and sample from a standard normal $\mathcal{N}(0, \mathbf{I})$ for the rest $M-1$ frames ($\{f^i_{T}\}_{i=2}^{M}$) in a video with $M$ frames. Our starting noise at $t=T$ can be thus written as a composition of reference and sampled noise $f_T = [x^r_{T},f^2_{T}, f^3_{T}, \dots]$. We note from \cref{sec: bck_diff} that, temporal attention and convolution layers within the T2V U-Net influence a frame with features from other frames during denoising. This means that naive denoising with starting noise $f_T = [x^r_{T},f^2_{T}, f^3_{T}, \dots]$ yields artefacts and inconsistent frames (\cref{fig:ablations}), as the reference frame is influenced by sampled noise at every timestep.

We begin by improving the joint denoising of $x^r_T$ with other frames by learning noise tokens $f^\text{train}_t = [f^2_t, \dots]$ for more accurate denoising of the first frame. Next, we modify self-attention maps in spatial and temporal attention units of $\epsilon_\theta$ to preserve identity of $I_s$ across all frames of generated animation. Similar to TF-ICON \cite{lu2023tf}, we inject self-attention and cross-attention maps between the reference $x^r_{T}$ and sampled $f^i_{T}$ noise during denoising (\cref{fig:model_1}).

\begin{figure*}
    \centering
    \vspace{-0.3cm}
    \includegraphics[width=\linewidth]{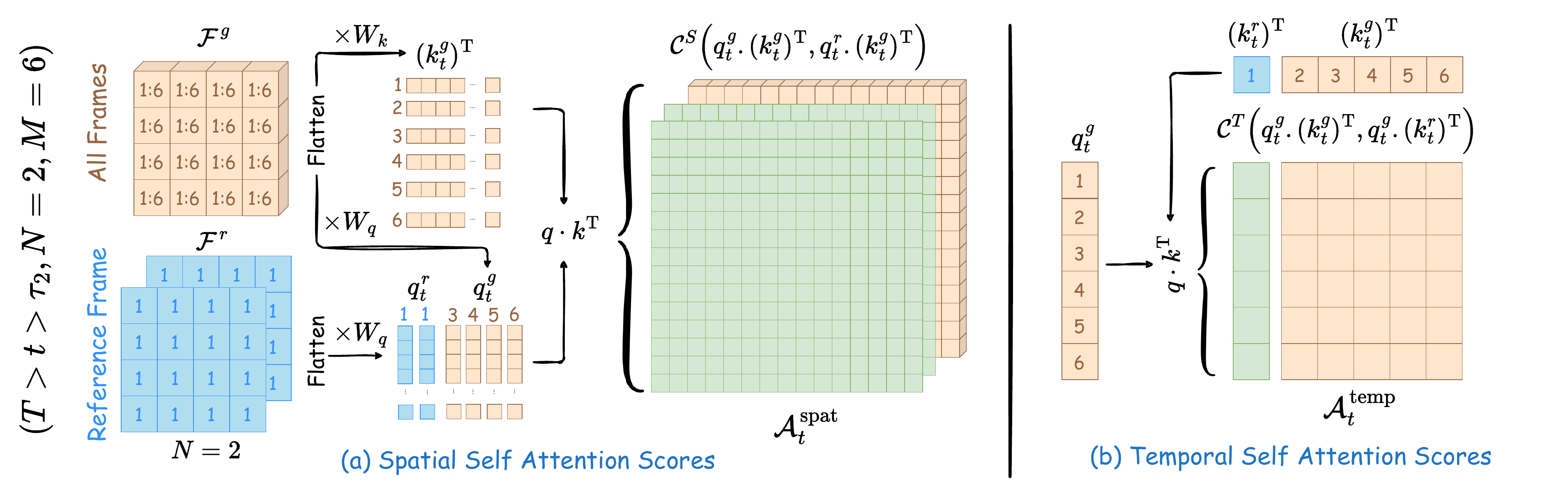}
    \vspace*{-1cm}
    \caption{We parallelly perform denoising of reference noise $x^r_t$ and that of all frames $f^i_t$. Query-key pairs from reference frame denoising ($q^r_t, k^r_t$) are used to influence video generation through cross-attention with ($q^g_t, k^g_t$).\vspace{-0.6cm}}
    \label{fig:attn}
\end{figure*}

\subsection{Iterative Frame Alignment}
\label{subsec: alignment}
At every timestep $t$ for  $T>t>\tau_1$, we perform iterative refinement \cite{chefer2023attend} of sampled noise $f^\text{train}_t = [f^2_t, \dots]$ to align with reference noise $x^r_t$. Since coarse features like object locality and attributes are denoised very early in the diffusion process \cite{chefer2023attend}, we restrict our refinement to the first few timesteps with a timestep threshold $\tau_1$. For timestep $t$ ($\in [T,\tau_1]$), we denoise the reference noise $x^r_{t}$ as $\eta_1 = \epsilon_\theta(x^r_{t}, t, \mathcal{P}_\text{null})$, using null prompt $\mathcal{P}_\text{null}$. Following the hypothesis that diffusion denoising signals can act as features \cite{tumanyan2023plug}, we use $\eta_1$ as the ground truth feature of the first frame. We then obtain the denoising signal for joint denoising of $x^r_t$ and $f^i_t$ at the current timestep $t$ as: 
\begin{equation}
\vspace{-0.1cm}
 [\eta'_i]_{i=1}^{M}  = \epsilon_\theta([x^r_{t},f^\text{train}_{t}], t, \mathcal{P}_\text{input})
\end{equation}
where $[\eta'_i]_{i=1}^{M}$ refers to the predicted denoising signal (and feature-map) for each of the $M$ frames for text prompt $\mathcal{P}_\text{input}$. We calculate the loss between the predicted feature map for the first frame and the ground truth feature map as 
$\mathcal{L}_\text{align} = || \eta'_1 - \eta_1 ||^2_2$. Since the predicted features ($\eta'_1$) were influenced both by $x^r_t$ and $f^{\text{train}}_t$ with temporal attention (\cref{sec: bck_diff}), we can backpropagate gradients from $\mathcal{L}_\text{align}$ to optimise $f^\text{train}_{t}$. The optimization of $f^\text{train}_{t}$ improves joint denoising of $x^r_t$ and $f^i_t$ at every timestep and significantly improves consistency of generated frames with the input frame (\cref{fig:ablations}). Iterative frame alignment doesn't take away stochasticity of generation, as we can still observe variation in generated videos with different starting noise $\{f^i_{T}\}_{i=2}^{M}$. 

\subsection{Guided Denoising with Attention Composition}
We guide the denoising of sampled noise $\{f^i_{t}\}_{i=2}^{M}$ with compositional attention maps at each timestep $T>t>\tau_2$. For this, we parallelly perform \textit{(i)} joint denoising of all frames $\epsilon_\theta([x^r_{t},f^\text{i}_{t}], t, \mathcal{P}_\text{input})$, and \textit{(ii)} denoising of reference frame only $\epsilon_\theta([x^r_{t}], t, \mathcal{P}_\text{null})$. Similar to frame alignment in \cref{subsec: alignment}, we guide denoising with attention composition only for early timesteps (threshold $\tau_2$), where the diffusion model is generating structurally significant information. 

We begin by obtaining the query-key pairs for computing spatial and temporal (\cref{sec: bck_diff}) self-attention from $\epsilon_\theta([x^r_{t}], t, \mathcal{P}_\text{null})$ with projection matrices $W_q$ and $W_k$ as 
\vspace{-0.2cm}
\begin{equation}
q^{r}_t = \mathcal{F}^r_t \cdot W_q, \; \; \; k^{r}_t = \mathcal{F}^r_t \cdot W_k    
\vspace{-0.2cm}
\end{equation}
where $(q^{r}_t, k^{r}_t)$ represents the query-key pair for the reference noise, calculated from intermediate feature maps $\mathcal{F}^r_t$ at every layer of U-Net $\epsilon_\theta$. Next, we obtain query-key pairs ($q^g_t$, $k^g_t$) from joint denoising $\epsilon_\theta([x^r_{t},f^\text{i}_{t}], t, \mathcal{P}_\text{input})$ as 
\vspace{-0.2cm}
\begin{equation}
    q^{g}_t = \mathcal{F}^g_t \cdot W_q, \; \; \; k^{g}_t = \mathcal{F}^g_t \cdot W_k
\vspace{-0.2cm}
\end{equation}

\noindent Spatial and Temporal self-attention scores for joint denoising of all frames are generally calculated with $q^{g}_t$, $k^{g}_t$ query-key pairs as $\mathcal{A}^g_t = q^{g}_t. (k^{g}_t)^\text{T} / \sqrt{d_\text{dim}} $. Instead, we compose these self-attention scores with cross-attention against reference query-key pairs ($q^{r}_t$, $k^{r}_t$). As discussed in \cite{lu2023tf}, self-attention maps in denoising diffusion frameworks hold significant semantic information. Composing spatial and temporal self attention maps helps preserve coarse-grained and fine-grained reference sketch features. By constructing both spatial and temporal self-attention as a composition with reference query-key pair ($q^{r}_t$, $k^{r}_t$), we can influence frame-generation with high correspondence to the input sketch $I_s$. We discuss specific details of these compositions below:

\noindent \textbf{Spatial Attention Composition:} Spatial attention is performed over the flattened $H_sW_s$ dimension for $H_s \times W_s$ size visual features, with query-key $(q^g_t,k^g_t) \in \mathbb{R}^{B_s\times(H_sW_s)\times D_s}$. $B_s$ attention maps of dimension $(H_sW_s)\times(H_sW_s)$ are constructed with $q.k^\text{T}$ based on feature similarities in query-key ($q,k$) pairs. Computing cross-attention between $q^r_t$ and $k^g_t$ (green blocks in \cref{fig:attn}a) for any $B_s$ here, helps connect spatial features of $q^r_t$ with matching features from $k^g_t$. This in turn helps impose reference features on generated frames. However, $q^r_t \in \mathbb{R}^{B'_s\times(H_sW_s)\times D_s}$ is a query from a single frame with $B'_s<B_s$. To influence all frames with $q^r_t$, we repeat the reference frame across $N$ frames as $\{x^r_T\}_{i=1}^N$ for $N>1$ obtaining multi-frame reference query (See \cref{fig:attn}a blue blocks). We set $N\sim M$ for early timesteps and reduce to $N=1$ with a linear schedule across $T>t>\tau_2$. This prevents generated frames from reducing to static $I_s$. For $N<M$ and $B's<B_s$, we complete the partially constructed spatial self-attention map with self-attention between $q^g_t$ and $k^g_t$ (almond coloured blocks in \cref{fig:attn}a). Following this attention composition $\mathcal{C}^S$ from \cref{fig:attn}a, the attention 
scores $\mathcal{A}^\text{spat}_t$ and maps $\mathbf{A}^\text{spat}_t$ are:

\vspace{-0.5cm}
\begin{align}
    \mathcal{A}^\text{spat}_t &= \mathcal{C}^S\Big(q^{g}_t. (k^{g}_t)^\text{T}, q^{r}_t. (k^{g}_t)^\text{T} \Big) / \sqrt{d_\text{dim}} \\ 
    \mathbf{A}^\text{spat}_t &= \text{Softmax}(\mathcal{A}^\text{spat}_t)
\end{align}
\vspace{-0.6cm}

\noindent \textbf{Temporal Attention Composition:} For temporal attention units, self-attention is performed on the temporal dimension (across frames), with ($q^{g}_t$, $k^{g}_t$) $\in \mathbb{R}^{B_t\times M\times D_t}$ for $M$ frames. 
Temporal self-attention helps distil features from different frames for the  generation of the current frame. Consider a single-feature query-key-value set $\{q,k,v\} \in \mathbb{R}^{B\times M\times1}$, for temporal self-attention in a video with $M$ frames. The attention map can be calculated as $\mathbf{A} = \text{Softmax}(q.k^\text{T}/\sqrt{d_\text{dim}})$, where $\{q_i.k_1, q_i.k_2, \dots,q_i.k_M \}$ determines the influences of value-features $\{v_1, v_2 \dots, v_M\}$ for composing $i^\text{th}$ frame. 
In our case, we directly control the influence of the first frame (value $v^1$) in generating other frames by replacing self-attention units (almond blocks in \cref{fig:attn}b) with cross-attention against the reference (green blocks in \cref{fig:attn}b). Specifically, we compute the cross between reference key $k^r_t$ (single frame, blue in \cref{fig:attn}b) and query $q^g_t$, filling the rest of attention scores $\mathcal{A}^\text{temp}_t $ and maps $\mathbf{A}^\text{temp}_t$ with $(q^g_t$,$k^g_t$) self-attention following $\mathcal{C}^T$ as:

\vspace{-0.4cm}
\begin{align}
    \mathcal{A}^\text{temp}_t &= \mathcal{C}^T\Big(q^{g}_t. (k^{g}_t)^\text{T}, q^{g}_t. (k^{r}_t)^\text{T} \Big) / \sqrt{d_\text{dim}} \\ 
    \mathbf{A}^\text{temp}_t &= \text{Softmax}(\mathcal{A}^\text{temp}_t)
\end{align}
\vspace{-0.5cm}

\noindent \textbf{Motion v/s Fidelity}: Aligned with findings in \cite{gal2024breathing}, we observe the motion-fidelity trade-off, where more dynamic animations often lead to loss of identity of the input sketch in later frames. We offer a mode of control over this trade-off, where parameter $\lambda$ controls motion in the generated video. We design this parameter through control over temporal self-attention composition by naive scaling of $k^r_t$ that increases the first frame influence (like $v_1$): 
\vspace{-0.2cm}
\begin{equation}
     k^r_t = k^r_t \cdot (1 +  \lambda \cdot 2e^{-2})
\vspace{-0.2cm}
\end{equation}
Lower $\lambda$ yields more motion while higher  $\lambda$ improves stability and resemblance to input sketch.

\begin{algorithm}

\caption{Sketch animation pipeline}
\begin{algorithmic}[1] 
    \State \textbf{Input:} Sketch $I_s$, prompt $\mathcal{P}_\text{input}$
    \State \textbf{Setup:} $x^r_T \leftarrow \text{Inv}(I_s)$ \Comment{Inversion}
    \Statex \hspace{2.8em}$\{f^i_{T}\}_{i=2}^{M} \sim \mathcal{N}(0, \mathbf{I})$ \Comment{Random Sampling}
    \For{$t \gets T$ to $0$}
    
        \If{$t \geq \tau_1$}
        \State $f^\text{train}_t = [f^2_t, \dots]$
            \For{$\mathcal{I} \gets 0$ to $\mathcal{I}_\text{max}$}
                
                \State $\eta'_1 = \epsilon_\theta([x^r_t, f^\text{train}_t], t, \mathcal{P}_\text{input})[0]$
                \State $\eta_1 = \epsilon_\theta(x^r_t, t, \mathcal{P}_\text{null})$
                \State $\mathcal{L}_\text{align} = || \eta'_1 - \eta_1||^2_2$
                \State $f^\text{train}_t \leftarrow f^\text{train}_t - \alpha_t \nabla_t \mathcal{L}_\text{align}$
            \EndFor
            \State $f^i_t \gets f^\text{train}_t$
        \EndIf
        \If{$t \geq \tau_2$}
        \State $N = \text{sched}(t, M)$ \Comment{Linear Schedule}
        \State $q^r_t,k^r_t = \epsilon_\theta([x^r_t]^N, t, \mathcal{P}_\text{null})$
        \State $\eta_{1:M} \gets \epsilon_\theta([x^r_t, f^i_t], t, \mathcal{C}^T, \mathcal{C}^S, q^r_t, k^r_t, \mathcal{P}_\text{input})$
        
        \Else
        \State $\eta_{1:M} \gets \epsilon_\theta([x^r_t, f^i_t], t, \mathcal{P}_\text{input})$
        
        \EndIf

    \State $[x^g_{t-1}, f^i_{t-1}] \gets \text{Denoise}(\eta_{1:M}, [x^g_{t}, f^i_{t}])$
        
    \EndFor
    \State \textbf{Return:} $[x^r_{0}, f^i_{0}]$
\end{algorithmic}
\end{algorithm}

\subsection{Implementation Details}

We construct a synthetic dataset of vector animations for text prompts using recent works in text to vector animation generation \cite{gal2024breathing}. We train a LoRA $\epsilon_\theta$ of the Modelscope T2V 3D UNet \cite{Wang2023ModelScopeTT} with a rank of $4$ for $2500$ iterations on these text-animation pairs. For generating raster sketch animations, we set the number of frames $M$ to $10$, and adjust $N$ as $N = \text{max}(1, (M-T-t))$. For longer videos with more frames (\cref{sec: ex}), we use the final frame of current video as the sketch input for a new video, stitching the videos together to extrapolate animation. We set thresholds $\tau_1 = \frac{2T}{5}$ and $\tau_2 = \frac{3T}{5}$ as fractions of the total timesteps $T=25$.

\begin{figure}[!htbp]
\hspace{-0.5cm}
\centering
\begin{floatrow}

    \begin{minipage}[t]{.45\linewidth}
    \vspace{-0.3cm}\hspace{-0.5cm}
        \begin{tikzpicture}
            \begin{axis}[
                legend cell align={left},
                legend style={font=\scriptsize,at={(0.25,0.8)},anchor=west},
                legend columns=1,
                legend style = {fill=none, font=\scriptsize,draw=none, cells={align=left}},
                scale only axis=true,
                width = 0.9\linewidth,
                height= 2.2cm,
                xmin=1.5, xmax=4.5,
                ymin=-1, ymax=6,
                xtick={2,2.772,3.466,4.159},
                xtick style={draw=none},
                xticklabels={$\text{log}_e(\text{strokes})$, 16, 32, 64},
                xticklabel style={rotate=0,font=\scriptsize},
                ytick= {0, 0.693, 1.607, 2.487, 3.401},
                yticklabels={1, 2, 5, 12, 30},
                grid style=dashed,
                axis y line=center,
                axis x line=bottom,
                ylabel={\scriptsize{$\text{log}_e(\text{minutes})$ on RTX 4090($\downarrow$})},
                ylabel style={at={(-0.1,1.2)}},
                ytick style={draw=none},
                yticklabel style={xshift=0.1cm, font=\scriptsize},
                extra x tick labels={Cut},
            ]                              
            \addplot[color= grey, mark = *, mark size=0.4mm, thin, smooth]
            coordinates{(2.079, 2.577) (2.772,2.671) (3.466,2.922) (3.871,3.157) (4.159,3.337)};
            \addlegendentry{Live-Sketch \cite{gal2024breathing}}
            \addplot[color= gold , mark = *, mark size=0.4mm, thin, smooth]
            coordinates{(2.079, -0.059) (2.772,-0.059) (3.466,-0.059) (3.871,-0.059) (4.159,-0.059)};
            \addlegendentry{Ours}
            \end{axis}
        \end{tikzpicture}
    \end{minipage}

  \begin{minipage}[t]{.45\linewidth}
      \vspace{-0.3cm}
      \begin{tikzpicture}
            \begin{axis}[
                legend cell align={left},
                legend style={font=\scriptsize,at={(0.25,0.8)},anchor=west},
                legend columns=1,
                legend style = {fill=none, font=\scriptsize,draw=none, cells={align=left}},
                scale only axis=true,
                width = 0.9\linewidth,
                height= 2.2cm,
                xmin=9, xmax=22,
                ymin=9, ymax=28,                      
                xtick={11, 15,17,20},
                xtick style={draw=none},
                xticklabels={No. of frames, 15, 17, 20},
                xticklabel style={rotate=0,font=\scriptsize},
                ytick= {10, 15, 20},
                yticklabels={10, 15, 20},
                grid style=dashed,
                axis y line=center,
                axis x line=bottom,
                ylabel={\scriptsize{Memory in GiB ($\downarrow$})},
                ylabel style={at={(-0.1,1.23)}},                
                ytick style={draw=none},
                yticklabel style={ xshift=0.1cm, font=\scriptsize},
                extra x tick labels={Cut},
            ]                              
            \addplot[color= grey, mark = *, mark size=0.4mm, thin, smooth]
            coordinates{(10, 14.38) (12,16.10) (15,18.03) (17,19.48) (20,21.62)};
            \addlegendentry{Live-Sketch \cite{gal2024breathing}}
            \addplot[color= gold , mark = *, mark size=0.4mm, thin, smooth]
            coordinates{(10, 10.025) (12,10.898) (15,11.910) (17,12.50) (20,14.59)};
            \addlegendentry{Ours}
            \end{axis}
        \end{tikzpicture}
  \end{minipage}
\end{floatrow}

\vspace{-0.3cm}
\caption{Time and compute needs of Live-Sketch \cite{gal2024breathing} and our method for increasing number of strokes and frames respectively.}
\label{fig: time}
\end{figure}
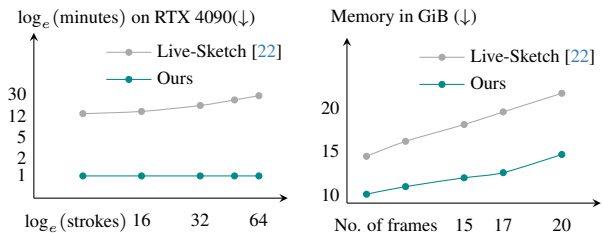

\begin{figure*}[!htbp]
\centering
\includegraphics[width=\linewidth]{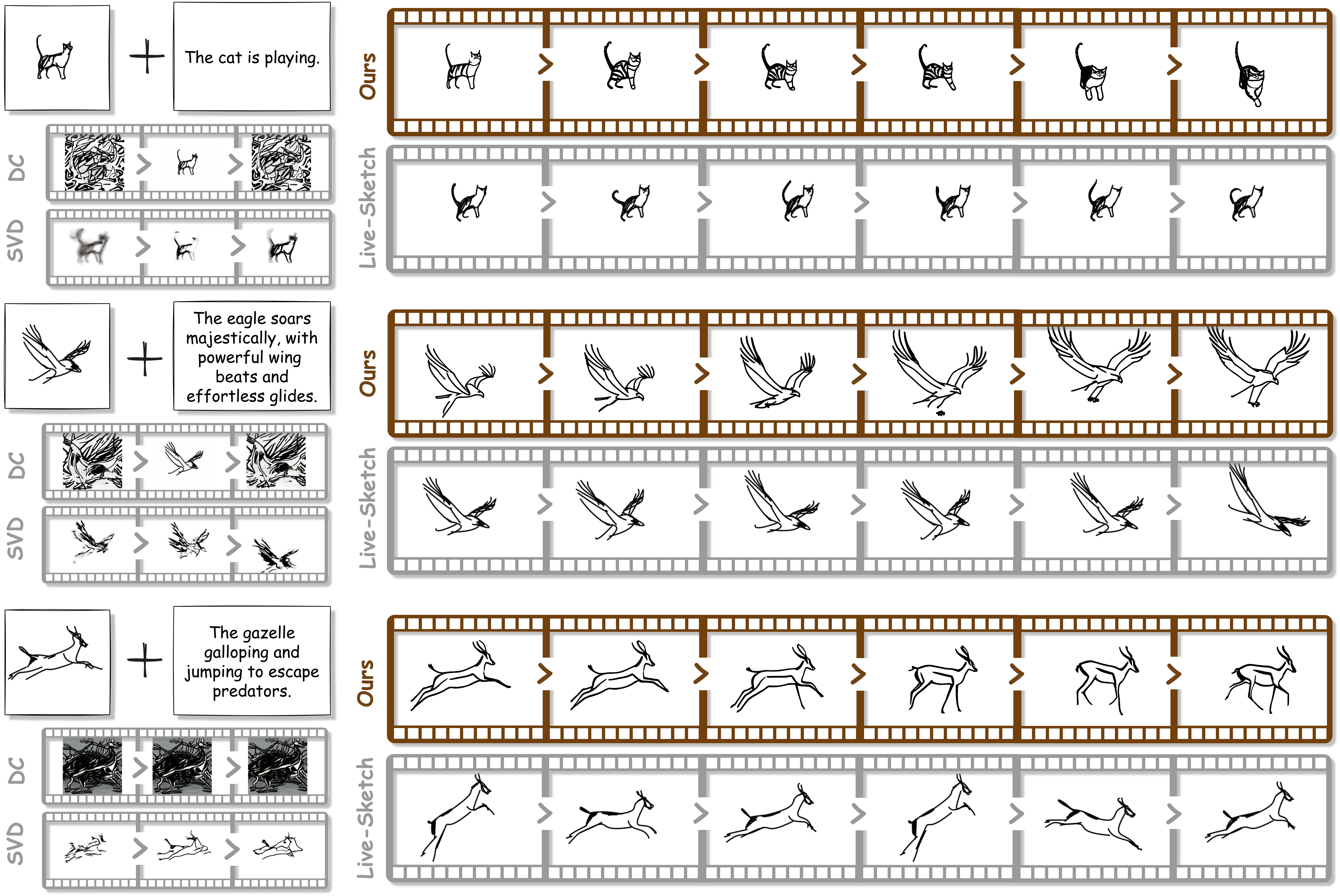}
\vspace*{-0.7cm}
\caption{Qualitative comparison of our method against vector animation algorithm Live-Sketch \cite{gal2024breathing} and raster video generation methods SVD \cite{blattmann2023stable} and DynamiCrafter (DC) \cite{xing2023dynamicrafter}. Live-Sketch \cite{gal2024breathing} preserves sketch identity by constraining local animations between vectors, but has limited motion capacity. SVD \cite{blattmann2023stable} and DC \cite{xing2023dynamicrafter} cannot preserve sketch identity, suffering from sketch-photo domain gap. Our method performs dynamic animations that align with text prompts, without losing sketch identity.\vspace*{-0.6cm}}
\label{fig:main_fig}
\end{figure*}

\section{Results and Comparisons}

\label{sec: ex}

We compare our results against recent works on video generation (\cref{fig:main_fig}, \cref{tab: quant-clip}, \cref{tab: quant-user}), notably \textbf{Live-Sketch} \cite{gal2024breathing} that generates reference-free sketch animations from input sketches and text prompts. \textit{Live-Sketch} animates vector sketches by obtaining a displacement field of vector coordinates through Score Distillation Sampling \cite{poole2022dreamfusion} (SDS) based optimisation. Besides being memory intensive and time consuming (\cref{fig: time}), \textit{Live-Sketch} suffers from pitfalls like limited motion and pre-defined number and arrangement of strokes. In addition, we note in \cref{fig: time} that \textit{Live-Sketch} performance depends on the number of strokes in sketch input, offering poor scaling with sketch complexity.

We also include comparisons against recent image-to-video (I2V) approaches like \textbf{DynamiCrafter} \cite{xing2023dynamicrafter} and \textbf{SVD} \cite{blattmann2023stable}. \textit{DynamiCrafter} conditions pre-trained T2V models on images by projecting them to a text-aligned representation space and using them to guide frame denoising. \textit{SVD} trains an image-to-video diffusion model based on pre-trained text-to-image (T2I) Stable Diffusion 2.1 \cite{rombach2022high}. We additionally compare with our fine-tuned text to sketch animation model (\textbf{T2V LoRA}) in ablative studies to directly analyse the influence of sketch prompts and performance without it.

Finally, we note from comparisons in \cite{gal2024breathing} that approaches involving skeletons \cite{smith2023method} expect sketches to be humanoid, failing to produce videos with generally inanimate objects like ``a plant in a pot", ``a cup on the table", or non-humanoid birds and animals. This prevents skeleton-based approaches to be useful for diverse sketch categories. For reference, we present qualitative comparisons with \textbf{Animated Drawings} \cite{smith2023method} in the Suppl.

\noindent \textbf{User Study}: We construct a user study, where we show videos generated by competitor methods and ablative configurations. Specifically, we compare with \textit{Live-Sketch}, base T2V LoRA, and our pipeline without attention composition. We ask each user to rank videos from all methods for a given text prompt in terms of \textit{(i)} faithfulness to text prompt and \textit{(ii)} consistency with input sketch. We convert the average ranks ($r$) to scores and normalize, as $\frac{t-r}{t}$ for comparing $t$ methods. Finally, we ask users to subjectively grade all videos (Mean Opinion Score) based on generation quality from $0$ (worst) to $1$ (best). We summarise the results of this study in \cref{tab: quant-user} with additional details in the Suppl.

\subsection{Text and Sketch to Video Generation}

We summarise our results of text+sketch to raster animation in \cref{fig:main_fig} and include qualitative comparisons against \textit{Live-Sketch} and I2V approaches \textit{SVD} and \textit{DynamiCrafter} (DC). We note that our animations are more flexible, offering new strokes and sketch configurations at each frame. This is particularly useful, for example, in animating for prompts like ``\texttt{the cat is playing}" where the subject changes orientation and direction in the 3D space. Despite our algorithm and \textit{Live-Sketch} sharing the same T2V base-model \cite{Wang2023ModelScopeTT}, we can extract more dynamic and realistic motion priors for sketch animations. In comparisons with other I2V approaches like \textit{DynamiCrafter} and \textit{SVD}, we note that these approaches generate noisy animations, suffering from the sketch-photo domain gap.

\begin{figure}[!htbp]
    \centering
    \includegraphics[width=\linewidth]{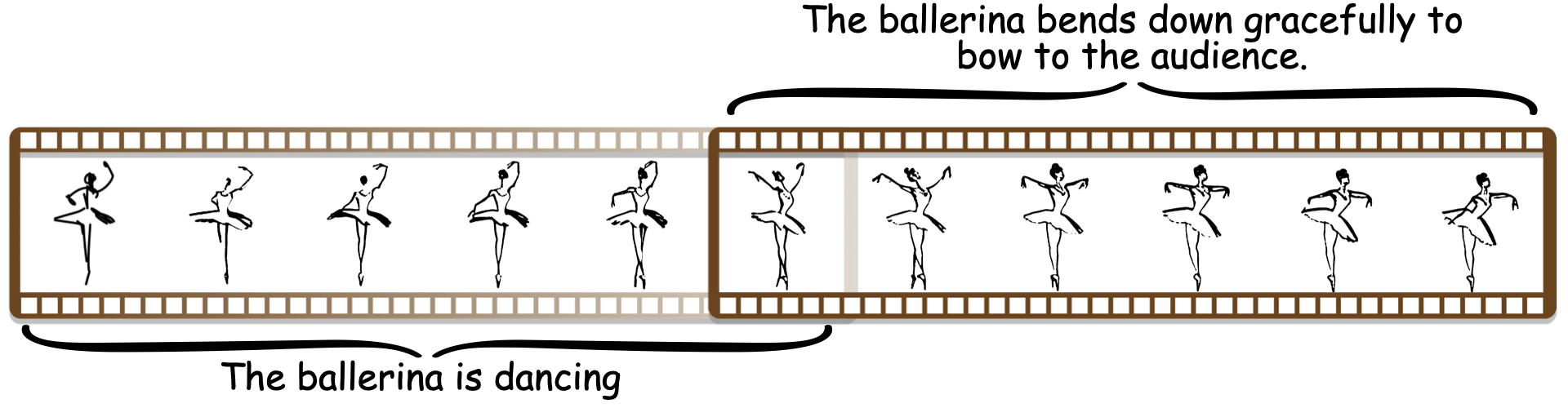}
    \vspace*{-0.8cm}
    \caption{Frame extrapolation allows us to construct complex animations by stitching multiple videos with different text prompts.}
    \label{fig:extrapolation}
\end{figure}

\noindent \textbf{Frame Extrapolation:} We demonstrate the construction of longer frame sequences with complex animation prompts in \cref{fig:extrapolation}. We break down complex actions into simple movements, generated using our text+sketch to animation pipeline. To preserve sketch identity across multiple animations, we use the last frame of one animation as the input sketch for next animation in the series. High consistency in generated frames helps preserve input sketch identity across multiple videos, while unconstrained raster animations allow performing complex actions without motion repetition.

In addition to qualitative analysis, we perform a quantitative study (\cref{tab: quant-clip}) similar to \cite{gal2024breathing}, using CLIP to measure metrics like ``sketch-to-video consistency" as the average similarity score between input and generated frames.  We also use X-CLIP \cite{ma2022x} to measure ``text-to-video alignment" as the similarity score between generated video and text prompt. We find that \textit{Live-Sketch} performs better in retaining the structure of the original sketch, as it heavily constrains motion. Our algorithm, however, significantly outperforms \textit{Live-Sketch} in text-to-video alignment, demonstrating better distillation of motion priors for animation.

\begin{figure}[!htbp]
\centering

\includegraphics[width=\linewidth]{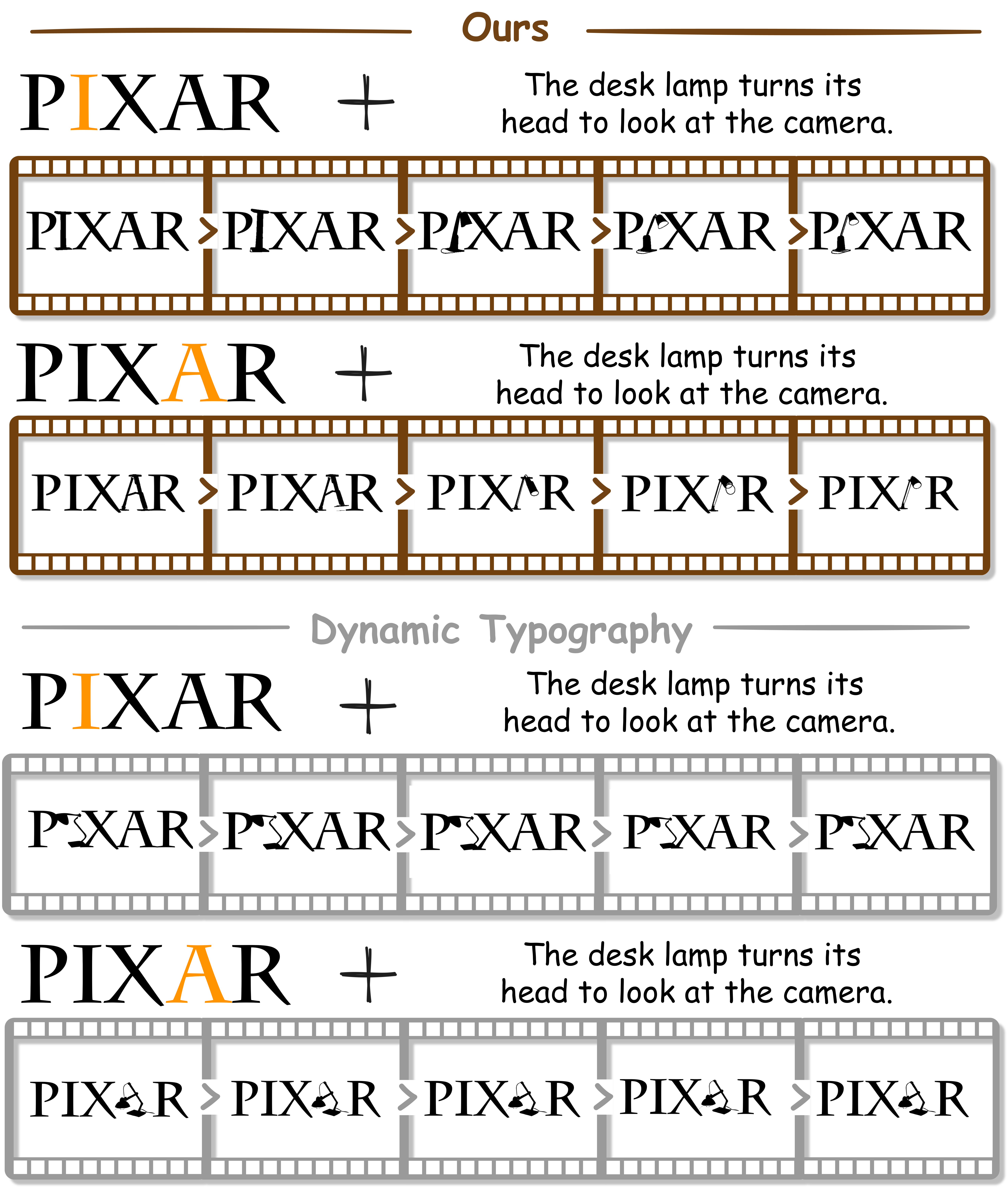}
\vspace*{-0.75cm}
\caption{ Qualitative comparison against vector animations from Dynamic Typography \cite{liu2024dynamic} for animating words with text prompts.\vspace{-0.4cm}}
\label{fig:typography}
\end{figure}

\begin{table}[ht]

    \centering
    \scriptsize
    
    \caption{ Comparing animations with CLIP-based metrics \\}

    \begin{tabular}{lcc}
        \toprule
        \textbf{Method} & \textbf{S2V Consistency ($\uparrow$)} & \textbf{T2V Alignment ($\uparrow$)} \\
        \midrule
        SVD \cite{blattmann2023stable} & 0.917 ± 0.004 & - \\
        T2V LoRA  & - & 0.158 ± 0.001 \\
        DynamiCrafter \cite{xing2023dynamicrafter}  & 0.780 ± 0.003 & 0.127 ± 0.003 \\
        Live-Sketch \cite{gal2024breathing} & 0.965 ± 0.003 & 0.142 ± 0.005 \\
        \midrule
        Ours  & 0.956 ± 0.004  & 0.172 ± 0.002 \\
        Ours @ $\lambda=0$ & 0.949 ± 0.002 & \textbf{0.174 ± 0.001} \\
        Ours @ $\lambda=1$ & \textbf{0.968 ± 0.003} & 0.170 ± 0.001 \\
        Ours w/o frame align & 0.952 ± 0.004 & 0.171 ± 0.001 \\
        Ours w/o $\mathcal{C}^T \& \; \mathcal{C}^S$ & 0.876 ± 0.004 & 0.168 ± 0.001 \\
        \bottomrule
    \end{tabular}
    \vspace*{-0.3cm}
    
    \label{tab: quant-clip}
    
\end{table}

\begin{figure}
    \centering
    
    \includegraphics[width=\linewidth]{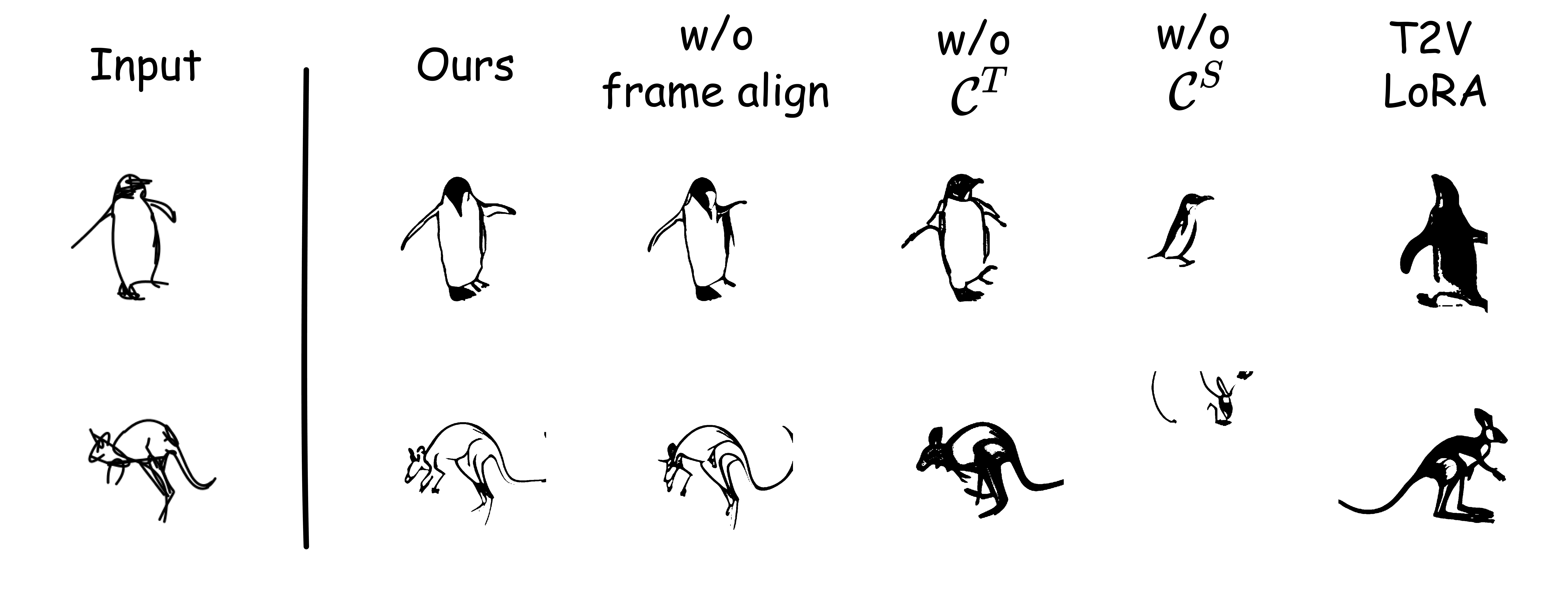}
    \vspace*{-0.95cm}
    \caption{Qualitative comparison of ablative configurations.}
    \label{fig:ablations}
\end{figure}

\begin{table}[ht]
\vspace{-0.1cm}
    \centering
    \scriptsize
    \begin{tabular}{lccc}

        \toprule
        \textbf{Method} & \textbf{Consistency ($\uparrow$)} & \textbf{Faithfulness ($\uparrow$)} & \textbf{MOS} ($\uparrow$) \\
        \midrule
        Live-Sketch \cite{gal2024breathing} & 0.51 & 0.44 & 0.63 \\
        T2V LoRA & 0.26 & 0.27 & 0.53 \\
        Ours  & \textbf{0.54} & \textbf{0.54} & \textbf{0.70} \\
        Ours w/o $\mathcal{C}^T \& \; \mathcal{C}^S$ & 0.20 & 0.25 & 0.43 \\
        \bottomrule
    \end{tabular}
    \vspace*{-0.2cm}
    \caption{Comparing animations with user study
    \\}
    \label{tab: quant-user}
\end{table}

\noindent \textbf{Ablative Studies:} We conduct ablative studies, where we qualitatively (\cref{fig:ablations}) and quantitatively (\cref{tab: quant-user} and \cref{tab: quant-clip}) analyse videos generated under different configurations by (i) changing hyperparameter $\lambda$ to re-balance the motion-fidelity trade-off in temporal self-attention $\mathbf{A}^{\text{temp}}_t$, (ii) removing attention composition for spatial $\mathcal{C}^{\text{S}}$ and temporal self-attentions $\mathcal{C}^{\text{T}}$, and 
(iii) removing frame alignment by skipping iterative refinement of sampled noise. In the Suppl., we perform additional ablations where we change hyperparameters $\tau_1$ and $\tau_2$ to demonstrate their effect on video quality and consistency. We note that lower $\lambda$ increases movement in frames (\textit{T2V alignment}) significantly while directly impacting sketch-to-video consistency in \cref{tab: quant-clip} ($\lambda$ ablative figures in Suppl.). At higher $\lambda$, we observe better fidelity to input sketches (\textit{S2V Consistency}) but more restricted motion (\textit{T2V alignment}). We also note that composing spatial and temporal self-attention is important for preserving coarse-grained and fine-grained sketch identity respectively (\cref{fig:ablations}). Finally, we observe that iterative refinement for frame aligning helps improve consistency in early frames by smoothing out fine-grained details.

\vspace{-2mm}
\subsection{Animating Words}
By removing temporal attention composition $\mathcal{C}^T$ completely, we can generate highly dynamic frames with reduced identity preservation. This helps us perform complex and visually rich animations with letters, where they morph from one form into another smoothly with diffusion motion priors. These complex morph animations allow for stylistic dynamic logo generation \cite{liu2024dynamic}. We compare with \textbf{Dynamic Typography} \cite{liu2024dynamic} the only other word animation model, that animates vector letters to resemble characters and objects. In \cref{fig:typography}, we demonstrate our animations to be unconstrained and have much more dynamic range than vector based \textit{Dynamic Typography} \cite{liu2024dynamic}. 

\begin{figure}
    \centering
    \includegraphics[width=\linewidth]{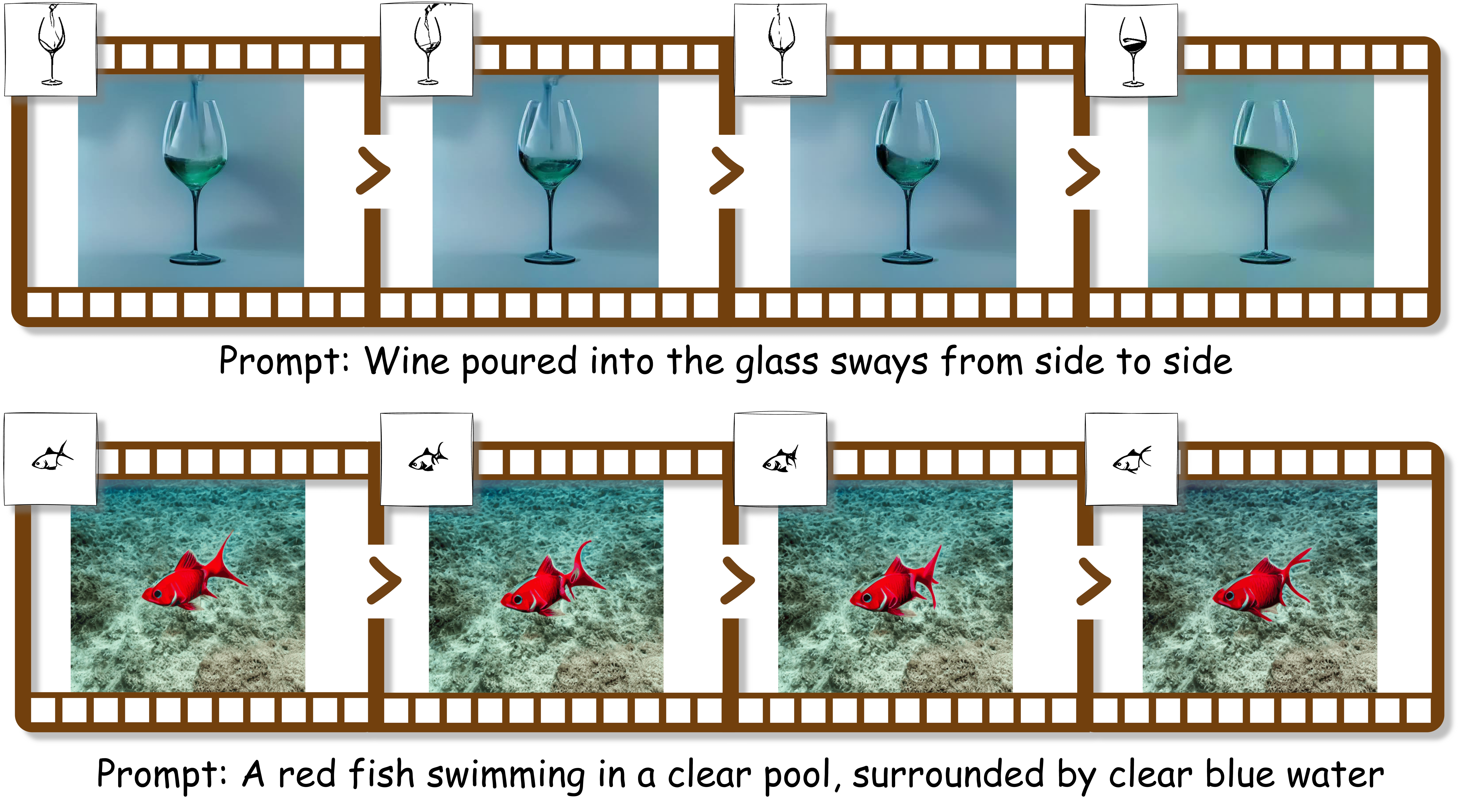}
    \vspace*{-0.8cm}
    \caption{We construct high resolution realistic videos using text prompts and skeleton-like guidance from generated raster frames.}
    \label{fig:real_vid}
\end{figure}

\subsection{Sketch Assisted Video Generation}
We demonstrate the applicability of sketch animations in real world video generation. By generating animation as object skeletons from sketch+prompt, we can assist the generation of real world videos (see \cref{fig:real_vid}). For this, we generate a sketch video from a text prompt and interpolate it with FILM \cite{reda2022film} to increase the frame-rate. We then convert the smooth sketch animation to a real world video by sketch-to-photo \cite{zhang2023adding,wang2024instantstyle} and diffusion-based video transfer \cite{ku2024anyv2v}, or, by directly using the sketch as a prompt for edge-map guided video generation \cite{chen2023control}.

\vspace{-0.2cm}
\section{Limitations}
\vspace{-0.2cm}
We note a primary limitation of our work in the slight stylistic resemblance of generated videos to CLIPasso \cite{vinker2022clipasso} sketches, owing to the uni-modality (in style) of our training data. Additionally, our pipeline handles sketch abstraction poorly, requiring high quality and geometrically accurate illustrations. Abstract inputs and irregular geometry often leads to the model correcting them on it's own from the first frame itself, resulting in poor sketch-to-video correlation.

Finally, our model is limited to the motion priors learned during pre-training of the ModelScope T2V network. Hence, while sketch animations are simpler than real world videos, we still face issues when generating motion in the form of extra limbs and inconsistent geometry.

\vspace{-0.2cm}
\section{Conclusion}
\vspace{-0.2cm}
We propose unconstrained raster sketches as potential alternatives to stroke-constrained vectors for sketch animations. Our raster animations are dynamic while preserving sketch identity, thanks to large-scale pre-training of T2V diffusion models. We outperform SOTA vector sketch animation in dynamic range and animation quality with both pre-trained vision-language metrics and user studies. Finally, we explore applications of our work for real world video generation, using recent work in diffusion-based video editing and spatial control for video generation.

{
    \small
    \bibliographystyle{ieeenat_fullname}
    \bibliography{main}
}

\end{document}